\documentclass[pre,onecolumn,showpacs,superscriptaddress,aps]{revtex4}

\usepackage{amsfonts,xfrac}
\usepackage{amsmath}
\usepackage{amssymb}
\usepackage{graphicx}
\usepackage{dcolumn}
\usepackage{dsfont}
\usepackage{times}
\usepackage{epsfig,multirow}
\usepackage[]{units}

\newcommand{\R}{{\mathbb R}}

\def\L{\mathcal{L}} 
\def\V{\mathcal{V}} 

\newcommand{\eps}{\varepsilon}
\newcommand{\OO}{\mathcal{O}}
\newcommand{\pa}{\partial}

\newcommand{\subfigimg}[3][,]{%
  \setbox1=\hbox{\includegraphics[#1]{#3}}
  \leavevmode\rlap{\usebox1}
  \rlap{\hspace*{5pt}\raisebox{\dimexpr\ht1-.5\baselineskip}{#2}}
  \phantom{\usebox1}
}

\begin{document}

\title{Exciting and Harvesting 
Vibrational States in Harmonically Driven Granular Chains}
\author{C. Chong\footnote{Email: cchong@bowdoin.edu}}
\affiliation{Department of Mechanical and Process Engineering (D-MAVT), \\ %
Swiss Federal Institute of Technology (ETH), 8092 Z\"urich, Switzerland}
\affiliation{Department of Mathematics, Bowdoin College, Brunswick, ME 04011, USA}
\author{E.  Kim }
\affiliation{Aeronautics \& Astronautics, University of Washington, Seattle, WA 98195-2400, USA}
\author{E. G. Charalampidis}
\affiliation{Department of Mathematics and Statistics, University of Massachusetts, Amherst, MA 01003-4515, USA}
\author{H.  Kim }
\affiliation{Aeronautics \& Astronautics, University of Washington, Seattle, WA 98195-2400, USA}
\author{F.~Li }
\affiliation{Aeronautics \& Astronautics, University of Washington, Seattle, WA 98195-2400, USA}
\author{P. G. Kevrekidis}
\affiliation{Department of Mathematics and Statistics, University of Massachusetts, Amherst, MA 01003-4515, USA}
\author{J.~Lydon}
\affiliation{Department of Mechanical and Process Engineering (D-MAVT), \\ %
Swiss Federal Institute of Technology (ETH), 8092 Z\"urich, Switzerland}
\author{C.~Daraio}
\affiliation{Department of Mechanical and Process Engineering (D-MAVT), \\ %
Swiss Federal Institute of Technology (ETH), 8092 Z\"urich, Switzerland}
\author{J.~Yang}
\affiliation{Aeronautics \& Astronautics, University of Washington, Seattle, WA 98195-2400, USA}
\date{\today}

\begin{abstract}
This article explores the excitation of different vibrational
 states in a spatially extended dynamical system through theory and experiment. 
As a prototypical example, we consider a one-dimensional packing of spherical particles (a so-called granular chain)
that is subject to harmonic boundary excitation. The combination of the multi-modal nature of the system and the strong coupling 
between the particles due to the nonlinear Hertzian contact force 
leads to broad regions in frequency where different vibrational 
states are possible. In certain parametric regions, we
demonstrate that the Nonlinear Schr\"odinger (NLS) equation predicts the
corresponding modes fairly well.
We propose that nonlinear multi-modal systems can be useful in vibration energy harvesting and discuss a prototypical framework for its
realization. The electromechanical model we derive predicts accurately the conversion from mechanical 
to electrical energy observed in the experiments.
\end{abstract}

\maketitle


\section{Introduction}\label{sec:intro}

Granular chains consist of closely packed arrays of particles that
interact elastically \cite{Nester2001}. The contact force can be tuned to yield near linear to purely nonlinear responses,
and the effective stiffness properties can also be easily changed by modifying the material,
geometry, or contact angle of the elements in contact \cite{Nester2001}. This remarkable tunability has made the topic
of granular chains an active research area over the past two decades (see the reviews \cite{Nester2001,sen08,theocharis_review}).
Granular chains have been proposed for numerous applications such as shock and energy absorbing 
layers~\cite{dar06,hong05,fernando,doney06}, actuating devices \cite{dev08}, 
acoustic lenses \cite{Spadoni}, acoustic diodes \cite{Nature11} and switches \cite{Li_switch}, as well as sound 
scramblers \cite{dar05,dar05b}. Examples of fundamental studies include solitary waves \cite{Nester2001,sen08,coste,pik} and 
dispersive shocks \cite{herbold,molin}. 

Time-periodic solutions have also been explored \cite{theocharis_review},
and constitute one of the main focal points of this article. Time-periodic solutions which are also localized in space (so-called breathers) have been explored in a host of nonlinear lattice models
during the 25 years  since their  theoretical inception, as has been 
summarized, e.g., in \cite{Flach2007}. 
Examples include optical waveguide arrays or photorefractive 
crystals~\cite{moti}, micromechanical cantilever arrays~\cite{sievers},
Josephson-junction ladders~\cite{alex},
layered antiferromagnetic crystals~\cite{lars3}, halide-bridged transition
metal complexes~\cite{swanson}, dynamical models of the DNA double 
strand \cite{Peybi} and Bose-Einstein condensates 
in optical lattices~\cite{Morsch}. In the context of granular chains, breathers have been studied in
 various settings, including monomers (chains where all particles are identical) \cite{dark}, chains with defects \cite{Theo2009},
 dimers (chains with a spatial periodicity of two) \cite{Theo2010} and trimers and quadrimers (chains
 with a spatial periodicity of three and four respectively) \cite{hooge13}. Gap effects
 related to transient behavior of driven dissipative granular chains have also been studied \cite{Herbold_Nesterenko_2009}.
 Time-periodic solutions of damped-driven granular chains (including breathers, but not exclusively)
 have been explored in monomers \cite{dark2}, defect chains \cite{Nature11}, dimers \cite{hooge12} and trimers \cite{Stathis}. 

Despite the large volume of studies
 of time-periodic solutions in granular chains, there has been very little attention paid to the high amplitude
 time-periodic solutions that exist in these models, especially from an experimental perspective. For example, a common theme
 in the studies of damped-driven granular chains is the transition between low-amplitude time-periodic states to quasi-periodicity or chaos \cite{Nature11,hooge12,Stathis}.
 The focus of this article will be on the different vibrational states in 
damped-driven granular chains.
Herein, 
we measure voltage variations related to different force transmission between particles
by embedding a piezoelectric sensor connected to a purely  resistive load in one of the beads of the chain. 
The resulting power is then calculated
 as $P=V^2/R$, where $V$ is measured the voltage drop over the resistor which has resistance $R$. 
 We are particularly interested in states that yield larger values of voltage (and hence power).
 We identify various vibrational states
 and characterize them through
experiments, numerical simulations and, in certain parametric regions, theory
for a large range of input (i.e. driving) frequencies. While fundamentally 
interesting in
 their own right, the spontaneous emergence of such states
under harmonically driven excitations could potentially
prove promising towards energy harvesting applications.
While systems incorporating nonlinearity or multiple modes
have been shown to harvest energy more efficiently than ones based on linear oscillators (see \cite{Erturk2011}, but also the books \cite{Erturk2013,Beeby} for a comprehensive treatment of energy harvesting), these two concepts have not yet been combined for the purpose of energy harvesting.
In this article we demonstrate that combination of nonlinearity and multiple modes of the granular chain lead to 
the possibility of high energy states for a broad range of input frequencies. In the future,  different systems, bearing resemblance to granular chains, could be
engineered to exploit similar nonlinear phenomena as the one described in this paper for use in energy harvesting applications.

The manuscript is organized as follows: in Secs.~\ref{sec:exp} and \ref{sec:theory} we describe the
experimental and theoretical set-ups respectively. The main results are presented in
Sec.~\ref{sec:main} where we demonstrate that the granular chain has the possibility for high energy states for a broad range of input frequencies.
In Sec.~\ref{sec:NLS} we provide an analytical prediction for the voltage production based on
a long-wave length approximation. We provide concluding remarks and future
directions in Sec.~\ref{sec:theend}.

\section{Experimental Setup} \label{sec:exp}

\begin{figure}
\begin{center}
\vspace{-0.2cm}
\mbox{\hspace{-0.7cm}
\includegraphics[height=.35\textheight, angle =0]{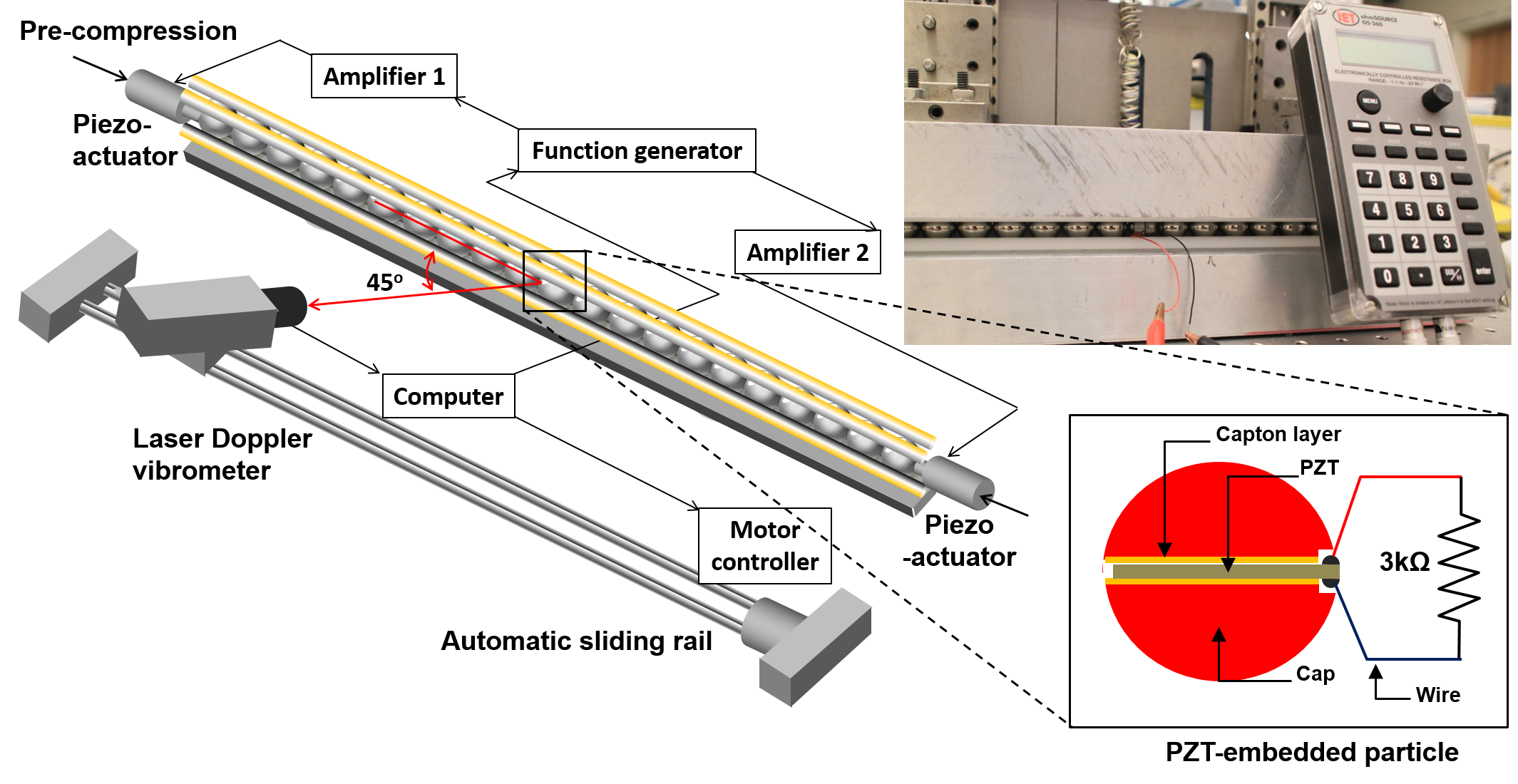}}
\end{center}
\caption{Schematic of the experimental set-up. The upper inset shows a digital image of the system, while the lower inset illustrates a schematic of the piezoelectric ceramic (PZT)-embedded sensor bead.}
 \label{fig:setup}
 \end{figure}

Figure~\ref{fig:setup} shows the experimental setup. The chain consists of 21 identical chrome steel spheres (see Table I for their mechanical properties and dimensions). These spheres are aligned using four polytetrafluoroethylene (PTFE) rods to constrain the particles' transverse motions while allowing their longitudinal vibrations with minimum friction against the supporting rods. The PTFE rods are mounted on the upper and lower stainless steel blocks to reduce the vibration effect of the rods (see the upper inset of Fig.~\ref{fig:setup}).

We apply 10 N force to the chain for static pre-compression using a spring and linear stage system~\cite{Li_PhysicaD_2013}. The chain is dynamically excited by two piezo actuators (Piezomechanik PSt 150/5/7 VS10) in contact with the first and last beads. A common excitation signal is generated using a function generator (Agilent 33220A), and it is sent to each actuator after being amplified through separate amplifiers (Piezomechanik LE 150/100 EBW) as shown in Fig.~\ref{fig:setup}. This enables us to excite both ends of the chain out-of-phase. 
 We measure the dynamic responses of the chain using a laser Doppler vibrometer (LDV, Polytec OFV-534). The vibrometer is placed in an automated sliding rail at a slanted angle (45$^{\circ}$), collecting the velocity profiles at specific particle spots. In this study, the localization and garnering of mechanical energy at the center of the chain is of particular interest. Therefore, besides the non-contact LDV method, we position a contact-based piezoelectric sensor bead at the center of the chain. For this, we fabricate a custom-made particle that embeds a lead zirconate titanate (PZT) disc between two halves of stainless steel spheres as shown in the lower inset of Fig.~\ref{fig:setup}~\cite{dar05}. The dimensions and properties of the PZT disc are in Table I. Note that the total mass and contact stiffness of the sensor bead are the same with that of the other beads, such that this bead can be treated as a regular particle in numerical simulations. The PZT disc is connected with an electronically controlled resistance box (IET OS-260), which is set to 3k$\Omega$ in this study. We measure the voltage applied to the resistor using an external oscilloscope, which can provide us with the information about the harvestable energy amount under simulated vibration conditions. 

To investigate its energy harvesting performance, we excite and test the granular chain in two different ways. One approach is that we excite the chain using a single frequency harmonic signal over a span of 0.1s. Given the speed of propagating waves in the chain, this is a period long enough to make the system reach a stationary state. We take the last 0.01s data to characterize the steady state response of the chain at specific frequencies. In this way, we use harmonic signals with the frequency from 6 kHz to 7.89 kHz at a 30 Hz interval. The other approach is that in a single round of testing, we dynamically sweep the frequencies from high (7.5 kHz) to low (6.5 kHz) ranges with 50 Hz step. Here, we run each frequency for 0.05s and take the last 0.01s data in each frequency interval. In both experiments, the amplitude of the excitation is consistent with $a=0.11 \mu m$. We perform each test three times to calculate their averages and standard deviations.


 %

\begin{table} \label{tab:params}
\begin{tabular}{ l c  c || l c  c  }
\multicolumn{3}{ c }{Mechanical Parameters} &
\multicolumn{3}{  c }{Electrical Parameters} \\
\hline 
Bead Mass &$M$ & $28.2$ g & Piezo Constant &$d_{33}$ & $360 \times 10^{-12}$  m/V \\ \hline
Bead Young's Modulus &$E$  &$200$ GPa  &Piezo Permittivity & $\epsilon^T$ &  $141664 \times 10^{-12}$ F/m \\ \hline
Bead Radius &$r$ & 9.53 mm   & Piezo Compliance &$S^E$& $72$ GPa$^{-1}$ \\ \hline
Bead Poisson's Ratio&$\nu$ &$0.3$ & Piezo Disc Area & $A$ & $283.5$ mm \\ \hline
Damping Coefficient &$\tau$ & 5 ms   &Piezo Thickness &$d$ & $0.3$ mm \\ \hline
 & &  & Resistance  & $R$ & $3$ $k\Omega$ 
 \\ \hline
\end{tabular}
\caption{Values of the electromechanical parameters. Mechanical properties are taken from nominal numbers available online and provided by vendors~\cite{efunda}, and geometrical dimensions are measured.}
\end{table}


\section{Theoretical Setup} \label{sec:theory}

In order to explore the theoretical model of interest for our system,
we briefly touch upon 
the granular chain~\cite{Nester2001}.  The model for the spatially homogeneous and damped-driven variant has the form,

\begin{equation} \label{eq:gc}
M \ddot{u}_n = \gamma[ \delta_{0} +   u_{n-1} -   u_{n}]_+^{3/2} - \gamma [ \delta_{0} +   u_{n} - u_{n+1}]_+^{3/2}  -  \frac{M}{\tau} \dot{u}_n ,  \quad n \in [1,N] \hspace{.2cm}\
\end{equation}
where $N$  is the number of beads in the chain, $u_n=u_n(t)\in\R$ is the displacement of the $n$-th bead from equilibrium position at time $t$, $\gamma =  \frac{ E \sqrt{2R} }{ 3(1-\nu^2)} $, is a material parameter, $M$ is the bead mass and $\delta_0$ is an
equilibrium displacement induced by a static load $F_0=\gamma \delta_{0}^{3/2}$. 
The bracket is defined by $[x]_+ = \mathrm{max}(0,x)$.
The form of dissipation is a dash-pot (with dissipation parameter $\tau$)
which can be interpreted as the friction between individual grains and PTFE rods. This form of dissipation
has been utilized in the context of granular chains in several previous works \cite{Nature11,dark2,Stathis}, but it is worth noting that
works such as \cite{herbold} (see also~\cite{vergara}) 
considered the internal friction caused by contact interaction between grains.
We consider out-of-phase harmonic boundary actuation,
\begin{equation} \label{eq:bc}
u_0 = a \cos( 2 \pi f_b t), \qquad u_{N+1} =   - a \cos( 2 \pi f_bt )
\end{equation}
where $a$  and $f_b$ represent the amplitude and frequency of the
actuation respectively.  
We insert a piezoelectric (PZT) sensor in the middle
bead since a voltage will be produced upon its deformation. 
In particular, a Lead-Zirconium-Titanate PZT sensor with area $A$ and thickness $d$ is used. 
To model the electromechanical system, 
we make use of the constitutive relations describing piezoelectric materials, which in the IEEE standard notation have the form~\cite{Preumont},
\begin{subequations}\label{eq:cont}
\begin{eqnarray} 
D &=& \epsilon^T E + d_{33} T, \\
S &=& d_{33} E + s^E T, 
\end{eqnarray}
\end{subequations}
where $D$ is the electric displacement, 
$E$ is the electric field, 
$T$ is the stress 
and $S$ is the strain. 
$\epsilon^T$ is the permittivity, 
$s^E$ is the inverse of the Young's modulus 
and $d_{33}$ is the piezoelectric constant. 
 For the parameters used in this study, see Table~I. 
 If we assume that the electrical and mechanical quantities are uniform throughout the PZT sensor 
 (with area $A$ and thickness $d$)
 then we infer $Q = DA$ where $Q$ is the total electric charge on the electrodes, $E = V/d$ where
 $V$ is the voltage between the electrodes and $f = A T$, where $f$ is the total force. Using these relations,
 we may rewrite Eq.~\eqref{eq:cont} as
 \begin{equation}\label{eq:cont_a}
\begin{pmatrix} Q \\ \Delta \end{pmatrix}
 = 
\begin{pmatrix}
C   & d_{33} \\ d_{33}  & 1/K_a
\end{pmatrix}
\begin{pmatrix} V \\ f \end{pmatrix}
\end{equation}
where $\Delta = Sd$ is the displacement (total extension) of the PZT, 
$K_a = \frac{A}{  s^E d}$ is the piezo stiffness  and $C = \frac{\epsilon^T A}{d}$ is the
piezo capacitance. Inverting Eq.~\eqref{eq:cont_a} yields,
\begin{equation}\label{eq:cont2}
\begin{pmatrix} V \\ f \end{pmatrix} = 
\frac{K_a}{C(1-k^2)}
\begin{pmatrix}
1/K_a   &- d_{33} \\ -d_{33}  &C
\end{pmatrix}
\begin{pmatrix} Q \\ \Delta \end{pmatrix}
\end{equation}
where $k^2 = \frac{d_{33}^2}{s^E \epsilon^T}$ is the electromechanical coupling factor.  
 
The equations of motion consisting of positions $q$ and velocities $\dot{q}$ of a mechanical system are the Euler-Lagrange equations of the Lagrangian $\L_m$ \cite{Arnold},
$$ 0  = \frac{d}{dt} \left(\frac{\pa \L_m}{\pa \dot{q}}\right) - \frac{\pa \L_m}{\pa q},$$
 with $ \L_m = T^*(\dot{q}) - \V(q) $ where $T^*$ and $\V$ are, respectively, the kinetic and potential energy of the mechanical system.
 This formalism can also be applied to electrical systems \cite{Preumont}. Using the flux-linkage coordinate $\lambda$ and voltage $\dot{\lambda}:=V$ 
 the equations of motion describing the electrical system are the Euler-Lagrange equations of the Lagrangian 
 $$ \L_e = W^*_e(\dot{\lambda}) -W_m(\lambda) $$
 where $W^*_e$ is called the electrical co-energy (which is analogous to the kinetic energy of a mechanical system)
 and $W_m$ is the magnetic potential (which is analogous to the potential energy of a mechanical system). For example,
 a capacitive element with capacitance $C$ in a circuit will be represented by the electrical co-energy $W^*_e(\dot{\lambda}) = \frac{1}{2}C\dot{\lambda}^2$.
 The electrical energy associated with the capacitive element $W_e(Q)$ is related to the co-energy through the Legendre
 transformation  $W_e(Q) = \dot{\lambda} Q - W^*_e(\dot{\lambda}) $, where $Q$ is the charge. Using the constitutive relation for
 a linear capacitor $C \dot{\lambda} = Q$ yields the energy stored in the capacitor $W_e(Q) = \frac{1}{2C}Q^2$. While
 this formulation is fairly trivial in the case of a stand-alone electrical system, it is quite natural when deriving
 the equations of motion for an electromechanical system. In this case, the Lagrangian will be the sum of the contributions from the mechanical system and the electrical system
 \cite{Preumont},
 $$ \L = T^* - \V + W^*_e - W_m.$$
 In what follows, we assume the corresponding magnetic potential energy $W_m$ is negligible, since e.g. we will not consider inductive elements.
  For a piezoelectric element the total power will be the sum of the electrical power $\dot{\lambda} \dot{Q}$ and mechanical power $ f \dot{\Delta}$. Thus, to compute the electric energy $W_e$, we have
 $$ \frac{d \, W_e}{dt} =   \dot{\lambda} \dot{Q}  +   f \dot{\Delta}. $$
 Upon substitution of $V=\dot{\lambda}$ and $f$ from Eq.~\eqref{eq:cont2} we see that the above is the total differential of
 $$  W_e = \frac{Q^2}{2C(1-k^2)} - \frac{d_{33} K_a}{C(1-k^2)} Q \Delta + \frac{K_a \Delta^2}{2(1-k^2)}. $$
 We can obtain the co-energy by making use of the Legendre transformation,
$$ W^*_e(\Delta,\dot{\lambda}) = \dot{\lambda} Q - W_e(\Delta,Q)  $$
which, after additional calculation, finally leads to,
\begin{equation} \label{eq:coe}
W_e^*(\Delta,\dot{\lambda})  = W_e^*(\Delta,V) =  C(1-k^2) \frac{V^2}{2} + d_{33} K_a V \Delta - K_a \frac{\Delta^2}{2}.
\end{equation}
We treat the bead with the embedded PZT as two half beads attached by a ``spring" (i.e. the PZT) \cite{Lydon},  where each half has mass $M/2$. We assume that these halves are located at sites $m$ and $m+1$. 
 Thus, the total extension of the PZT is $\Delta = \delta_1 + u_{m+1} - u_m$, where $\delta_1$ is the initial extension of the PZT due to any static force. With these assumptions, the Lagrangian of the electromechanical system has the form,
 $$ \L =  \sum_{n}  \frac{1}{2} M_n \dot{u}_n^2 +  \sum_{n \neq m} \left[ - \gamma \frac{5}{2} [ \delta_{0} +   u_{n} - u_{n+1}]_+^{5/2}  \right] + W^*_e(\Delta,V)$$
 where $W^*_e$ is defined by~\eqref{eq:coe} and $M_n=M/2$ for $n=m, m+1$ and $M_n = M$ otherwise. 
The equations of motion can be obtained starting from this Lagrangian:
\begin{subequations}\label{eq:EM}
\begin{eqnarray} 
M \ddot{u}_n &=& \gamma [ \delta_{0} +   u_{n-1} -   u_{n}]_+^{3/2} - \gamma [ \delta_{0} +   u_{n} - u_{n+1}]_+^{3/2}  -  \frac{M}{ \tau} \dot{u}_n ,   \quad  n \notin \{ m,m+1\}    \hspace{.2cm}   \label{eq:EMa} \\  
\frac{M}{2} \ddot{u}_m &= &\gamma [ \delta_{0} +   u_{m-1} -   u_{m}]_+^{3/2}  - K_a(\delta_1 +  u_m - u_{m+1} )  - d_{33} K_a V -   \frac{M}{2 \tau} \dot{u}_m ,    \hspace{.2cm}   \label{eq:EMb} \\  
\frac{M}{2} \ddot{u}_{m+1} &=& K_a( \delta_1 + u_m - u_{m+1} ) -  \gamma [ \delta_{0} +   u_{m+1} -   u_{m+2}]_+^{3/2}   + d_{33} K_a V -   \frac{M}{2 \tau} \dot{u}_{m+1} ,    \hspace{.2cm}   \label{eq:EMc} \\  
C(1 - k^2) \, \dot{V}  &=& -d_{33} K_a  ( \dot{u}_m - \dot{u}_{m+1}  ) -  \frac{V}{R} \label{eq:EMd}
\end{eqnarray}
\end{subequations}
where $F_0 = K_a \delta_1$.
The non-conservative terms appearing above are phenomenological terms added to account for the presence of damping. 
We model mechanical dissipation as a dash-pot (i.e. the terms $\frac{M}{\tau} \dot{u}_m,  \frac{M}{2 \tau} \dot{u}_m, \frac{M}{2 \tau} \dot{u}_{m+1} $ in Eqs.~\eqref{eq:EMa}-\eqref{eq:EMc} respectively). To convert voltage to power, we connect the electrodes with a purely resistive load $R$, which results in the non-conservative term
$V/R$ in Eq.~\eqref{eq:EMd}. Finally, we assume that the chain is finite in length, where the boundaries are given by Eq.~\eqref{eq:bc}.

 It is also worth mentioning that many studies on granular chains use embedded PZT elements to deduce the force applied on particles in a granular chain \cite{Nester2001}.
Indeed, it is only recently that the use of laser vibrometry has been introduced to probe the dynamics of the granular chain \cite{JKlaser}. Typically, the voltage response
measured from the PZT is calibrated empirically in order to obtain a force. Equation~\eqref{eq:EM} provides a more precise description of how the voltage
relates to the applied force on a given particle. Moreover, it describes how the electrical system can affect the dynamics of the mechanical system, and that
for large voltage responses, the dynamics can be severely altered. Although this aspect is outside the scope of the present article, it would be an interesting direction
for future studies.

\subsection{Linear Regime}

For dynamic displacements satisfying $\frac{|u_{n}-u_{n+1}|}{\delta_{0}}\ll 1$ we can expand the Hertzian forcing term in a Taylor series,
 \begin{equation} \label{eq:Taylor}
   \gamma[ \delta_0 + y]^{3/2} \approx \gamma \delta_0^{3/2} + K_2 y + K_3 y^2 + K_4 y^3, \qquad  K_2 = \frac{3}{2}\gamma \delta_0^{1/2}, \quad K_3 = \gamma \frac{3}{8}
\delta_0^{-1/2},\quad K_4 = -\gamma \frac{3}{48}\delta_0^{-3/2}.
\end{equation}
Thus, the linear equations (where $K_3=K_4=0$) have the form,
\begin{subequations} \label{eq:linear}
\begin{eqnarray} 
M_n \ddot{u}_n  &= & \beta_{n} (u_{n-1} - u_{n} )  - \beta_{n+1}(u_n - u_{n+1} )  -   \rho_n V -   \frac{M_n}{ \tau} \dot{u}_n ,    \hspace{.2cm}   \label{eq:lineara} \\  
R C(1 - k^2) \, \dot{V}  &=& -d_{33} K_a  R( \dot{u}_m - \dot{u}_{m+1}  ) -  V   \label{eq:linearb}
\end{eqnarray}
\end{subequations}
where,
\begin{equation} 
\beta_n = \left\{ 
\begin{array}{ll}
K_2 & n \neq m \\
K_a & n = m
\end{array}
\right.
\qquad
M_n = \left\{ 
\begin{array}{ll}
M & n \neq m,m+1 \\
M/2 & \mathrm{otherwise}
\end{array}
\right.
\qquad
\rho_n = \left\{ 
\begin{array}{cl}
d_{33} K_a  & n = m \\
-d_{33} K_a  & n = m+1 \\
0&  \mathrm{otherwise}
\end{array}
\right..
\end{equation}
The linear resonances of this system will be determined by Eq.~\eqref{eq:lineara}, which we view
as a homogeneous mass-spring system with two adjacent defects located at sites $m$ and $m+1$. Ignoring the mechanical and electrical damping
terms, one can show that there is a continuous (discrete) band of modes for an infinite (finite) system $f \in [0, f_0]$ where $f_0 = \sqrt{K_2/M}/ \pi$ is the cut-off
frequency of the homogeneous chain. The presence of the defects introduces two additional modes, which correspond
to the defects being in- or out-of-phase. They can be approximated respectively as \cite{Man2012},
\begin{eqnarray}
f^{(1)}_{defect} &\approx& \frac{1}{2 \pi}\sqrt{ \frac{1}{M^2}(s_1 \pm \sqrt{s_2})} \\
f^{(2)}_{defect} &\approx& \frac{1}{2 \pi}\sqrt{ \frac{1}{M^2}(s_3 \pm \sqrt{s_4})} 
\end{eqnarray}
where
$s_1 = 2 K_2 M$, $s_2 =  2  K_2^2 M^2$,
$s_3 = s_1 + 2 K_a M$ and $s_4 =  -4 [ 2 K_2 K_a + K_2( 2 K_a + K_2) ]M^2/2 + 4[K_a M + K_2 M]^2$. 
In-phase motion of the two defects will result in no voltage production (see Eq.~\eqref{eq:EMd}),
and thus $f^{(1)}_{defect}$ will not be relevant for our purposes. On the other hand, 
since $K_2 \ll K_a$,  out-of-phase motion of the defects will occur for frequencies
much larger than the maximum frequency obtainable in experiments  
(in the case of the parameters
of Table~I 
 the largest frequency is $f^{(2)}_{defect} \approx 349.63$ kHz, whereas $ f_0 = 7.765$ kHz).
Thus, since these defect modes  (either due to their nature
or due to their frequency range) cannot produce significant voltage
in the considered granular configuration, 
we resort to exciting 
several lower frequency (global) modes in order to produce
a voltage response. 

Considering once again the electrical and mechanical damping, and an external harmonic
excitation, we seek to compute the linear steady-state response. We use
the ansatz $u_n = \phi_n e^{i 2 \pi f_b t} $ and $V = \phi_v e^{i 2 \pi f_b t}$ 
and define $u_{0} = - u_{N+1} = a e^{i 2 \pi f_b}$, where $N$ is the number of nodes
(corresponding to $N-2$ beads and two bead halves) and where $a$ is the excitation
amplitude and $f_b$ is the excitation frequency. We solve the resulting
system of complex algebraic equations for $\{\phi_n,\phi_v\}$, where
the steady-state solution is then the real part of $\phi_n e^{i 2 \pi f_b t}$ 
 and $\phi_v e^{i 2 \pi f_b t}$.
Thus, if the natural response of the host structure is known, the material properties of the particles 
can in principle be selected such that the resonant peaks match the natural response. In a chain of cylindrical particles in contact, it has been demonstrated
recently that the stiffness $K_2$ can be easily tuned dynamically by alternating
the contact angle of cylinders \cite{alpha}. This could also be useful for the purpose of a linear based energy harvesting device where the
resonant response is tuned dynamically,
although we do not pursue that direction in this article. Rather we will exploit the nonlinearity to broaden the effective frequency response of the
system.




\section{Main Results} \label{sec:main}

 \begin{figure}
     \centering
   \begin{tabular}{@{}p{0.4\linewidth}@{\quad}p{0.4\linewidth}@{\quad}@{}}
    \subfigimg[width=\linewidth]{\bf (a)}{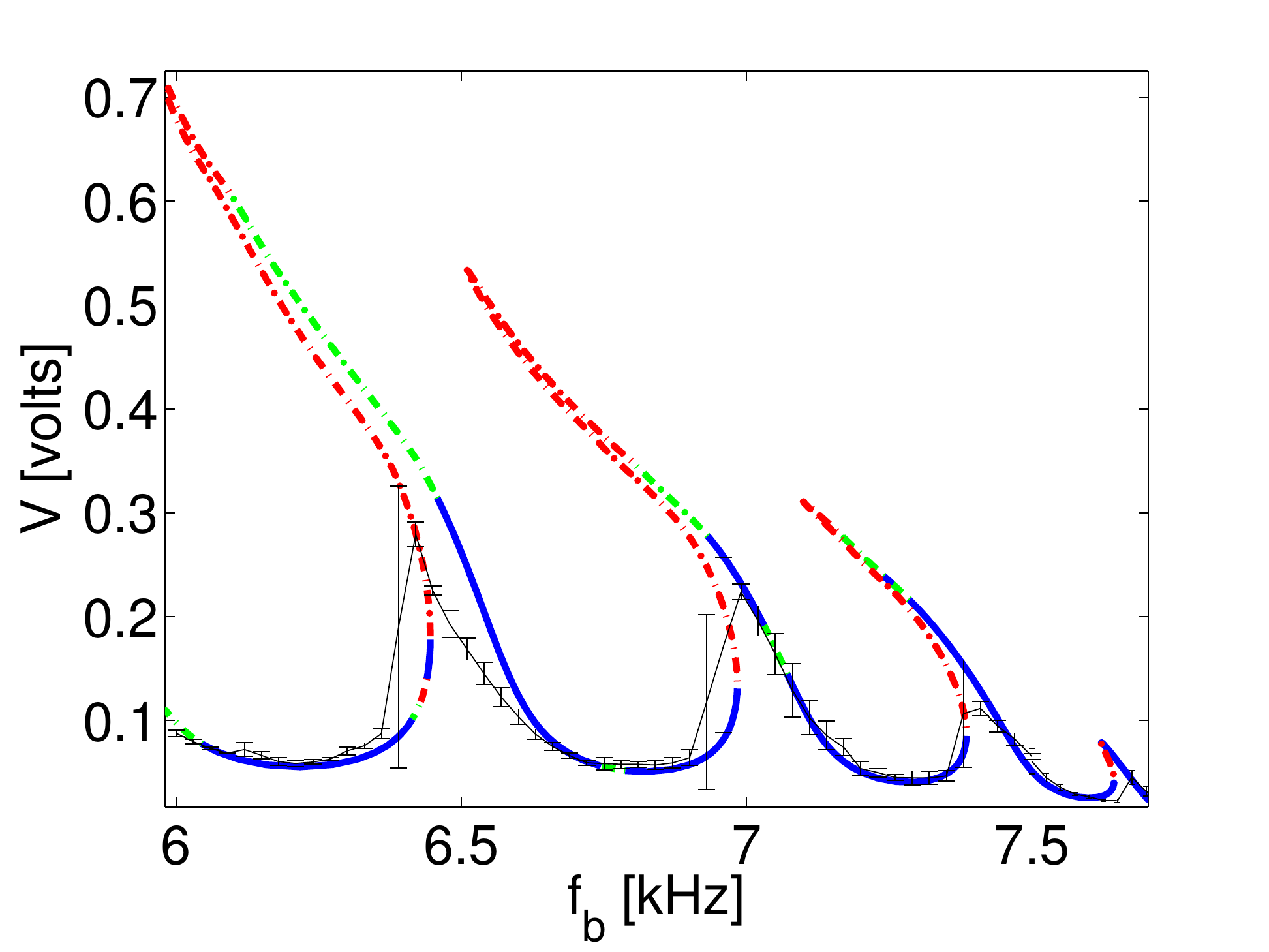} & 
     \subfigimg[width=\linewidth]{\bf (b)}{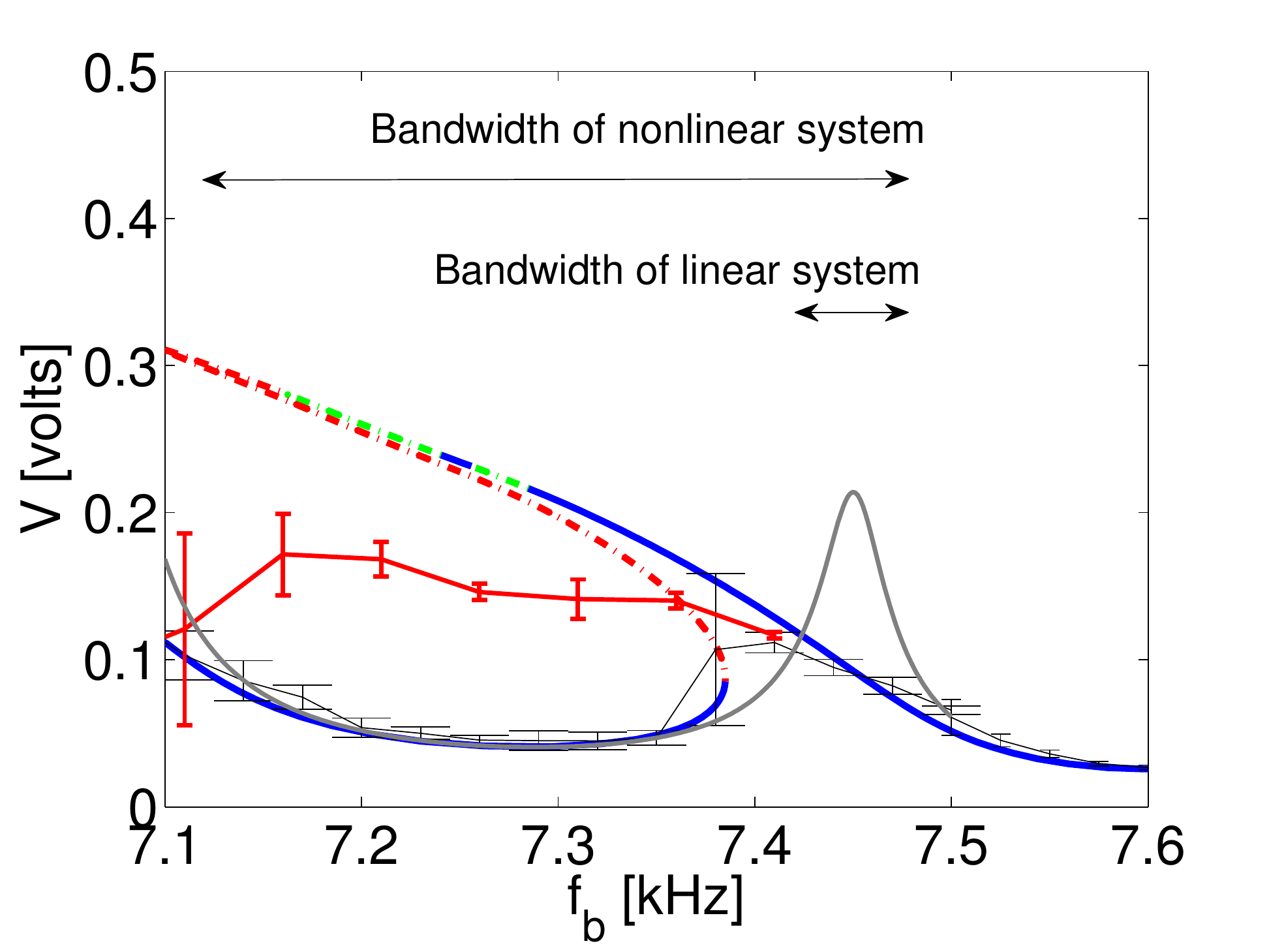}  
  \end{tabular}
      \subfigimg[width=.4 \linewidth]{\bf (c)}{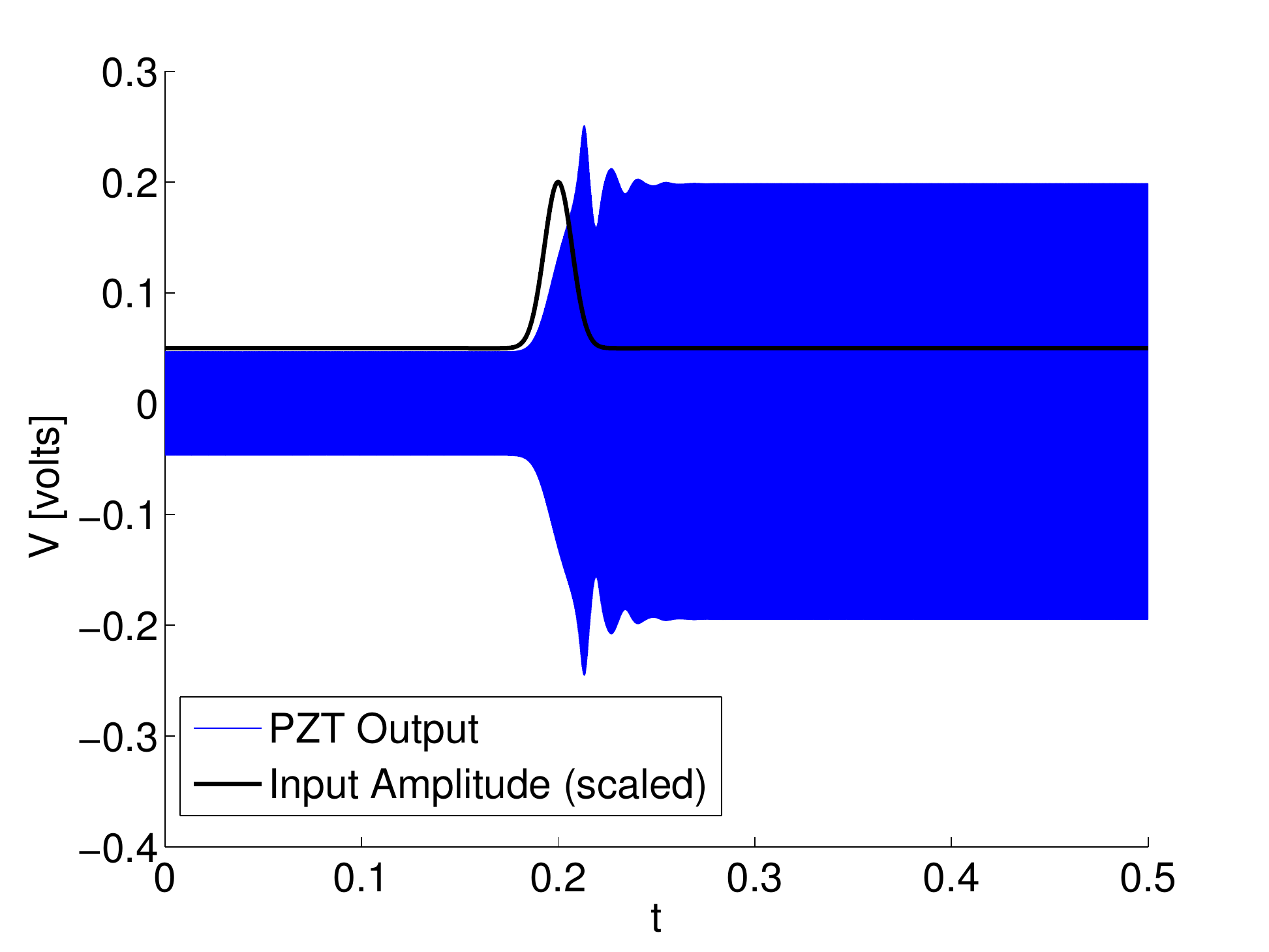}  
 \caption{(Color online) \textbf{(a)} Voltage amplitude of time-periodic solutions of Eq.~\eqref{eq:EM} vs driving frequency and a fixed driving amplitude of $a=0.11 \mu m$ (blue, red and green lines). 
 The color is associated with the stability properties
of the respective solutions (see details in the text). The experimental voltages shown (black markers with error bars) represent the average amplitude of the voltage over the final 10 ms of a 100 ms run. Each experimental run was initially at
 rest. Three experiments per marker were performed to obtain statistics.  \textbf{(b)} Zoom of panel (a) in the 7.1-7.6 kHz region. The thick red error bars correspond to an experiment where
 the actuation frequency was decreased dynamically (every $.05$ s). The values shown represent the average voltage amplitude over the final $.01$ s
 before the frequency was updated. The linear response is also shown (light gray line). \textbf{(c)} Voltage response for an amplitude of $a=0.11 \mu m$ and $f_b = 7.32$ kHz with a short amplitude burst where the maximum amplitude is increased by a factor of 1.5 (solid black line).
 }
 \label{fig:voltage_bif}
\end{figure}



 Since we are driving the chain harmonically, natural solutions to look for are time-periodic ones. We compute a solution with
period $T_b = 1/f_b$ with high precision by finding roots of the map $\mathbf{x}(T_b) - \mathbf{x}(0)$, where $\mathbf{x}(T_b)$ is 
the solution of Eq.~\eqref{eq:EM} at time $T_b$ with initial condition $\mathbf{x}(0)$.  Roots of this map (and hence
time-periodic solutions of Eq.~\eqref{eq:EM}) are found via Newton iterations.  An initial guess for the Newton
iterations can be provided by the steady-state solution of the linearized equations~\eqref{eq:linear}. Once a solution
is found a parametric continuation is performed.  In this study, we keep all parameters fixed, and vary
either the excitation amplitude $a$ or the excitation frequency $f_b$. For example, the solid lines in Fig.~\ref{fig:voltage_bif}(a)
are found by performing a pseudo-arclength continuation in the frequency $f_b$ of the time-periodic solutions. 
Spectral stability of these solutions is determined through the computation of Floquet multipliers corresponding to the Hill's equation that results from linearizing the equations of motion about a time-periodic solution \cite{Flach2007}.
A solution where all Floquet multipliers have modulus less than unity is asymptotically stable (solid blue lines in Fig.~\ref{fig:voltage_bif}(a)), marginally stable if the module of all multipliers is equal to unity,  and unstable
otherwise.  We classify the instabilities as either real if the
associated Floquet multiplier is purely real (dashed red lines in Fig.~\ref{fig:voltage_bif}(a)), or oscillatory if it has real and imaginary parts (dash-dot green lines in Fig.~\ref{fig:voltage_bif}(a)).
We make this distinction since (a) oscillatory instabilities are typically smaller in magnitude (see for example \cite{dark2,Stathis})
and (b) oscillatory instabilities suggest the presence of Neimark-Sacker bifurcations and hence of quasi-periodic
solutions.

As a first step, we validate the proposed model by comparing the above described time-periodic solutions of Eq.~\eqref{eq:EM} to experimental results.
Starting from the zero state (i.e. when the chain is at rest), the chain is driven at the boundaries with the excitation amplitude and frequency
being fixed.  We let the experiment run for $100$ ms and then compute the average amplitude of the voltage
over the final $10$ ms. If there is a stable periodic solution nearby, the $100$ ms is typically long enough
to allow the solution to reach that solution.  We use the theoretically predicted values of all parameters
with the exception of the precompression ($\delta_0 = 13.7$), which is used to match the linear resonant frequencies of the 
experiment and theory. The compliance of the PZT $S^E = 36 \mathrm{Gpa}^{-1}$ and the electromechanical coupling parameter $k = 0.87$
are modified to obtain better agreement of the voltage output.  The experimental measurements match the theoretical predictions
quite well in regions of asymptotic stability, see Fig.~\ref{fig:voltage_bif}(a), which validates
the model in Eq.~\eqref{eq:EM}. In Fig.~\ref{fig:voltage_bif} the voltage is shown since this is the quantity directly
measured from the experiments, however, the power can be computed using the formula $P = V^2/R$
where in this study $R=3$ k$\Omega$.

\begin{figure}[!ph]
\begin{center}
\vspace{-0.2cm}
\mbox{\hspace{-0.7cm}
\includegraphics[height=.35\textheight, angle =0]{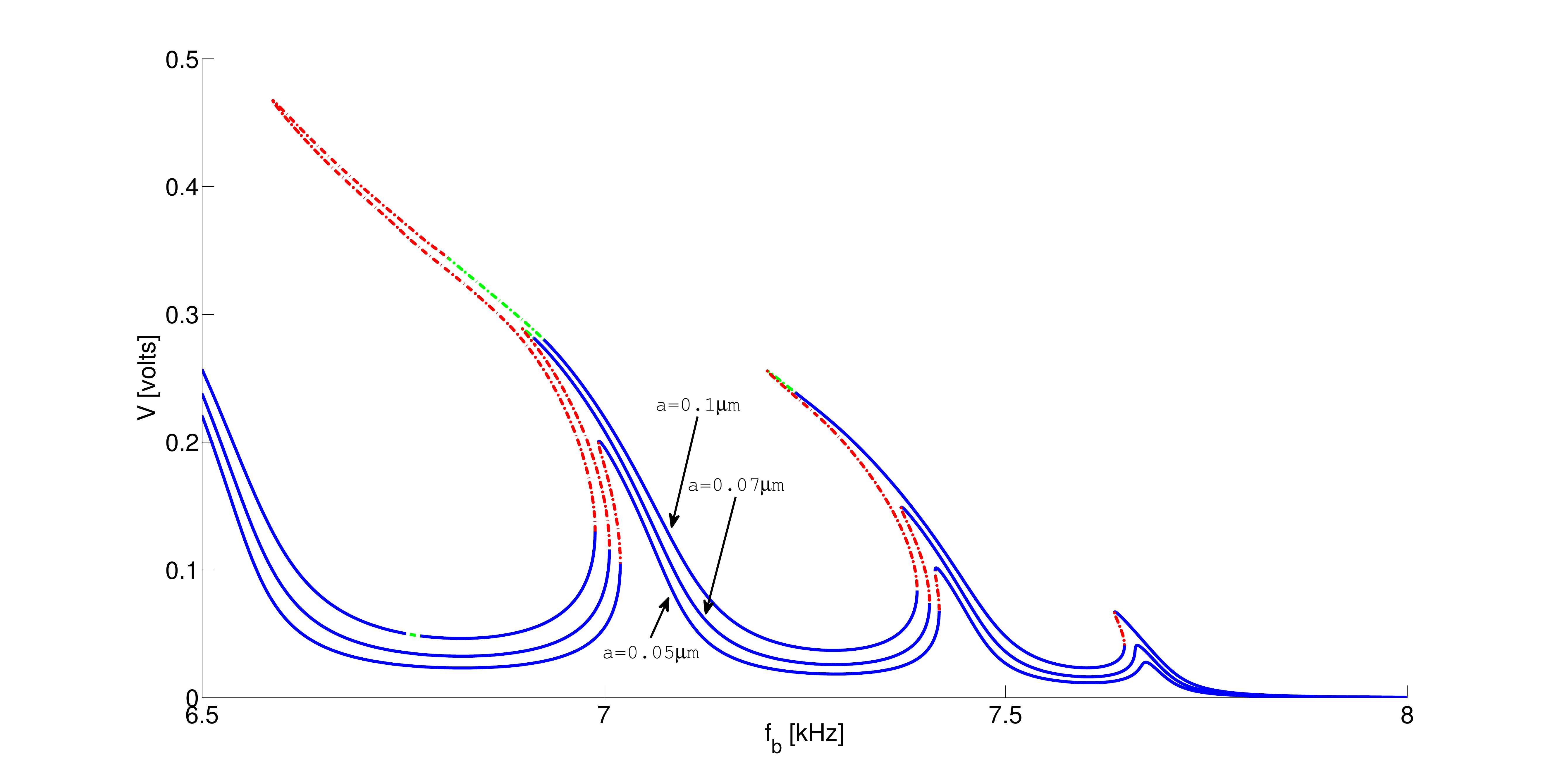}
}
\end{center}
\caption{(Color online) Bifurcation diagrams corresponding to the voltage
as a function of the actuation frequency $f_{b}$ and for various values of
the actuation amplitude of $a=0.05\,\textrm{$\mu$m}$, $a=0.07\,\textrm{$\mu$m}$
and $a=0.1\,\textrm{$\mu$m}$ are presented with smooth curves and increasing
order (see the arrows). Note that the blue segments correspond to stable 
parametric regions while the red and green ones correspond to real and oscillatory
unstable parametric regions, respectively (see text).}
\label{fig:voltage_bif_various_amps}
\end{figure}

The effect of the nonlinearity in the system is apparent in Fig.~\ref{fig:voltage_bif}(a), where the resonant peaks
have bent (in this case toward lower frequencies), such that there are high amplitude states for a 
large range of frequencies.  The bending trend of the resonant peaks as the driving
amplitude varies is shown in Fig.~\ref{fig:voltage_bif_various_amps}, see also \cite{Lydon}.
This, in conjunction with the multi-modal nature of our system,
implies the possibility of high amplitude states for a broad range of frequencies. However, if the system starts from the zero state, and the excitation
conditions are fixed,  the low amplitude state (which is typically stable) will be approached. See for example the
experimental measurements in Fig.~\ref{fig:voltage_bif}(a). However, if the excitation
frequency is changing dynamically (in this case from high to low), it is possible to access the high amplitude states. 
In Fig.~\ref{fig:voltage_bif}(b) experimental measurements are shown starting from the zero state as black
markers with error bars and experimental measurements that have a frequency that changes dynamically from high
to low frequency are shown as
thick red markers with error bars.  In the latter, one sees that  higher amplitudes have been achieved, although over the times considered in our experiment the nature of the resulting state (and its eventual 
asymptotics) has not yet been clarified/determined. It
is worth noting that, even though the high amplitude states may be 
unstable in certain regions of the frequency,
the corresponding instabilities are rather weak. Thus, over
transient dynamical regimes, it is possible to be near these large
amplitude states for relatively long times.  Another possibility to transition into a high amplitude
state is to apply a short amplitude burst to the excitation, see Fig.~\ref{fig:voltage_bif}(c). 
This aspect could be particularly useful for energy harvesting applications where the host structure
is subject to occasional large vibrations, and is worthy of future investigation in its own right.

A useful benchmark for this system is its linearized counterpart. Imagine that one can construct a linear system that has a resonant frequency
near the dominant frequency of some vibrating host structure, and that the parameters are tuned such that the voltage
production at full width at half maximum (FWHM)  is sufficient for the particular device the harvester is powering.
In Fig.~\ref{fig:voltage_bif}(b) the linear prediction is shown as the light gray line. If the host structure
vibrates at a resonant frequency (i.e. $7.45$kHz), then both the linear system and our nonlinear system would meet the voltage production
requirements.  If the excitation frequency differs
by about $.05$ kHz from the linear resonant frequency (due to e.g. mistuning), and the system is always starting from zero,
then our nonlinear system and the linear system will both fail to meet the minimum requirement. If in addition the excitation frequency is 
changing dynamically, then the nonlinear system has the potential to meet the minimum voltage requirement, whereas the linear system would fail.
For example, the FWHM of the linear response is approximately
 $0.1$ volts with a corresponding bandwidth of $0.05$ kHz. The effective bandwidth of the nonlinear response in the window
 shown in Fig.~\ref{fig:voltage_bif}(b) with voltages exceeding the FWHM ($0.1$ volts) is $0.3$ kHz. This represents a 500\% increase in bandwidth. For the chosen system size ($N=22$), there
are 10 resonant peaks located at  $\{1.1026,    2.1887,    3.2257,    4.1997,    5.0826,    5.8674,    6.5330,    7.0656,    7.4510, 7.6831\}$
kHz, each of which bends toward lower frequencies. The qualitative features shown in Fig.~\ref{fig:voltage_bif}(b) can be found around each of these frequencies implying the possible broadband capabilities of this system. 
While the introduction of additional modes can increase the bandwidth of the system,  it will also enhance the effect
of damping. 
This could be compensated by e.g. introducing multiple sensors throughout the chain to increase the power output.

%
%

\section{NLS approximation} \label{sec:NLS}

 \begin{figure} 
     \centering
   \begin{tabular}{@{}p{0.44\linewidth}@{\quad}p{0.4\linewidth}@{}}
    \subfigimg[width=\linewidth]{\bf (a)}{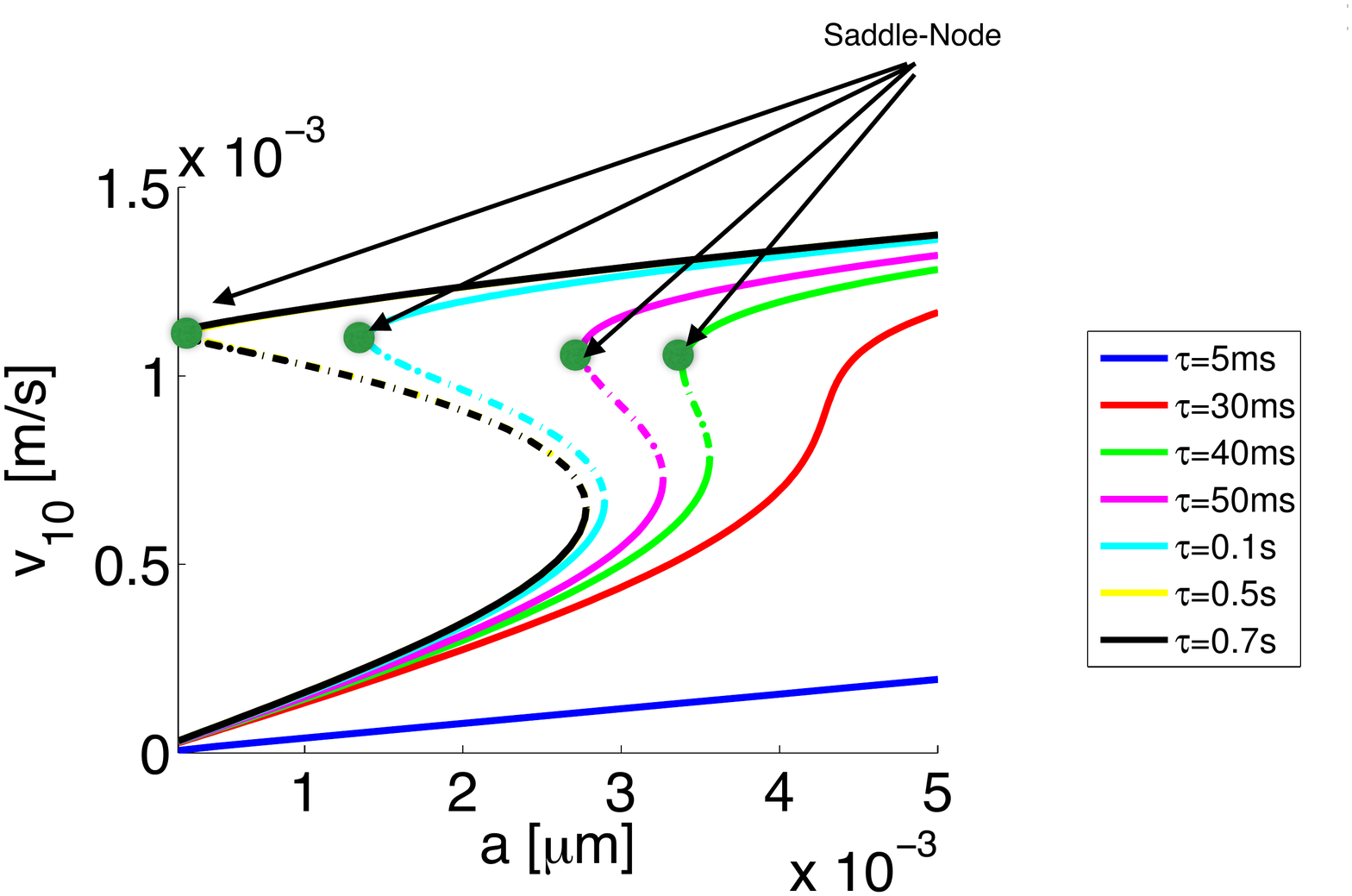} &
      \subfigimg[width=\linewidth]{\bf (b)}{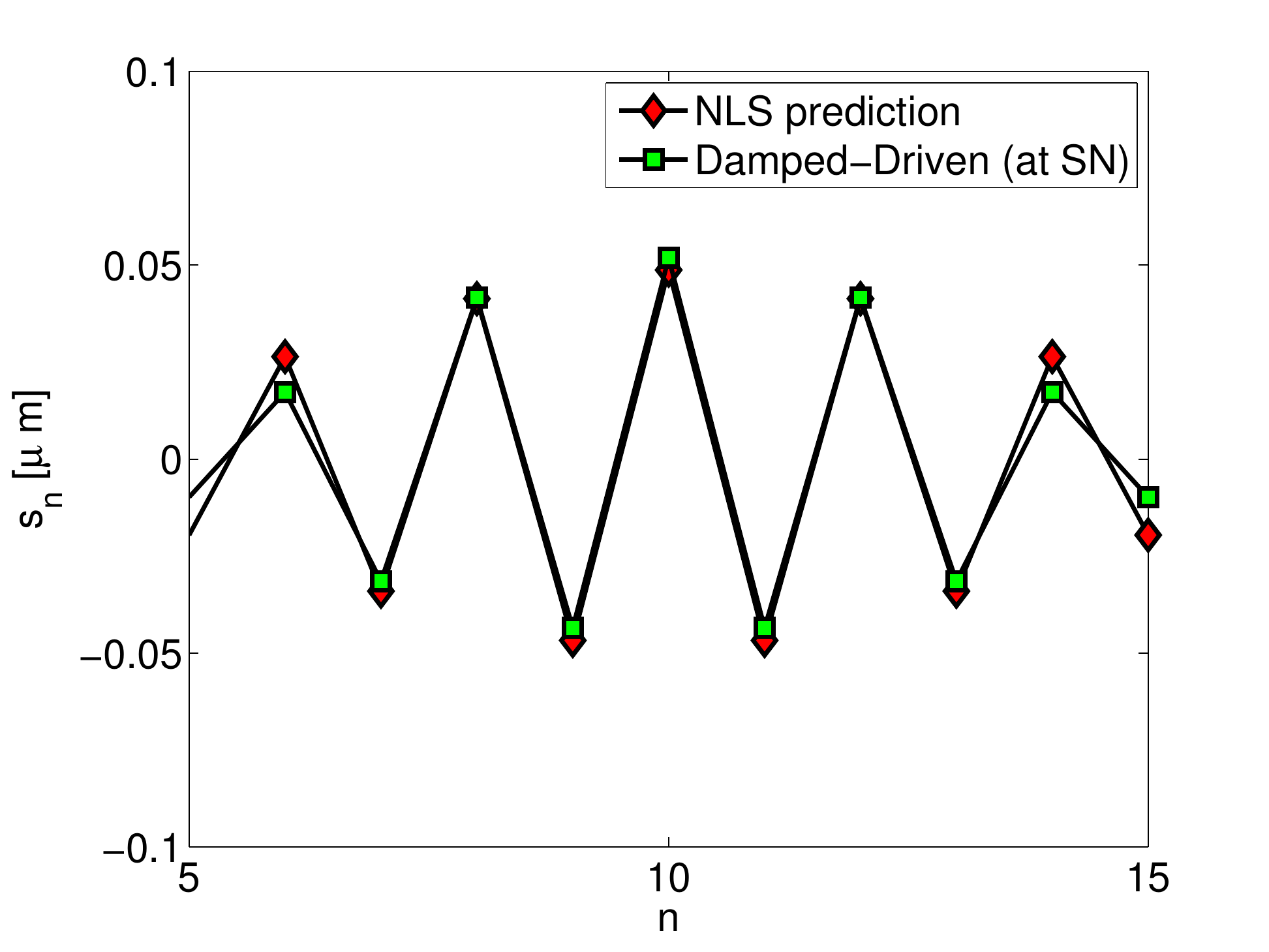} 
  \end{tabular}
   \subfigimg[width=.4 \linewidth]{\bf (c)}{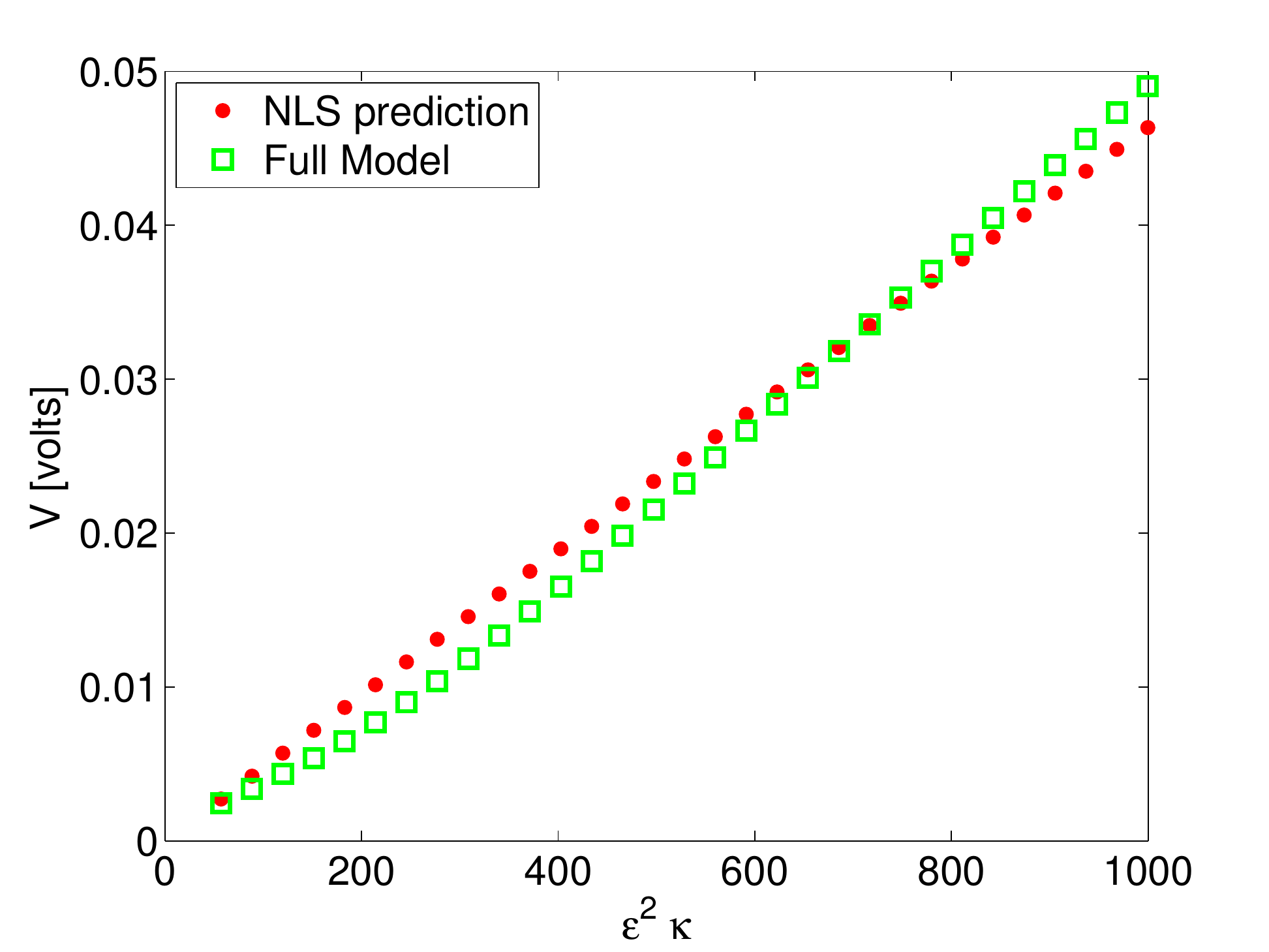} 
 \caption{(Color online) \textbf{(a)} Continuation in driving amplitude $a$ of time-periodic solutions of Eq.~\eqref{eq:gc} for a fixed driving frequency of $f_b = 7.29$ kHz and various
 dissipation values. The maximum velocity of 10th particle is shown. The saddle-node (SN) bifurcations which progressively approach 
the NLS limit as $a \rightarrow 0$ are labeled by the arrows. \textbf{(b)}  Comparison of the NLS approximation (red diamonds) and corresponding time-periodic solution of Eq.~\eqref{eq:EM} (green square) with a low dissipation constant $\tau = 0.7$s 
 \textbf{(c)} Steady-state amplitude of the voltage as predicted by Eq.~\eqref{eq:Vmax} (red circles) and the actual values given through Eq.~\eqref{eq:EM} (green squares) using
 the experimentally determined value of the dissipation $\tau = 0.5$ ms. }
 \label{fig:NLS}
\end{figure}

For certain parameter values, the Hamiltonian limit of Eq.~\eqref{eq:gc} (where $ \tau \rightarrow 0$ and $a \rightarrow 0$) can be useful towards 
understanding the full electromechanical
system of Eq.~\eqref{eq:EM}. This connection can be most easily seen by performing a continuation in driving amplitude $a$ and fixing all other parameters.
In Fig.~\ref{fig:NLS}(a) an amplitude continuation is shown for various values of the damping coefficient $\tau$. For small values of $a$ and $\tau$ one can see a low amplitude
state, (the near linear state which is ``dictated'' by the actuators) and two high amplitude nonlinear states that meet and collide in a saddle-node bifurcation
at some critical value of the driving amplitude $a_{cr}$; see the arrows of Fig.~\ref{fig:NLS}(a) and also~\cite{hooge12}.  Since $a_{cr} \ll 1$, the solution at the saddle-node is ``as close as possible" to the corresponding Hamiltonian 
solution.  In that same limit  
(of $ \tau \rightarrow 0$ and $a \rightarrow 0$)
the nonlinear Schr\"odinger (NLS) equation can be derived from Eq.~\eqref{eq:gc}
written in terms of the negative strain variable $y_n = u_{n-1} - u_{n}$, \cite{dark,Huang}. In particular, one defines the multiple-scale ansatz,
\begin{equation} \label{ansatz_strain}
y_n(t) \approx \psi_n(t) :=  \eps A(X,T) e^{i( k_0 n + \omega_0 t)}  + \mathrm{c.c.} + \mathrm{h.o.t}  \,,  \quad X=\eps( n + c t), \quad T = \eps^2 t,
\end{equation}
where $\eps \ll 1$ is a small parameter, effectively parameterizing the
solution amplitude (and also its inverse width). The 
substitution of this ansatz into the equations of motion
and equation of the various orders of $\epsilon$ leads to the dispersion relation 
$\omega_0 = \omega(k_0) :=  2 \sqrt{K_2/M}\sin(k_0/2) $, 
the group velocity relation $c = \omega'(k_0)$, 
and the nonlinear Schr\"odinger equation,
\begin{equation}\label{NLS}
 i \pa_T A(X,T) + \nu_2 \pa_X^2 A(X,T) + \nu_3 A(X,T)|A(X,T)|^2 = 0,
\end{equation}
where $\nu_2 = -\omega''(k_0)/2 > 0$ and 
\begin{equation} \label{gamma}
\nu_3 = \frac{K_3^2}{K_2^2}\tilde{\gamma} + \frac{3 K_4}{2 K_2} \omega(k_0) < 0,
\end{equation}
while
\begin{eqnarray} \label{gamma1}
\tilde{\gamma} & = & \frac{\omega(k_0)}{2} \left(\frac{\omega(2 k_0)}{2 \omega(k_0)-\omega(2k_0) }
-\frac{\omega(2 k_0)}{2 \omega(k_0)+\omega(2k_0) } \right.
\\ && \left. \qquad \qquad +
\frac{2\omega'(0)}{\omega'(k_0)-\omega'(0) }
-\frac{2\omega'(0)}{\omega'(k_0)+\omega'(0) }
 \right). \nonumber
\end{eqnarray}
The NLS equation is an integrable equation and has several exact solutions, such as the bright or (the relevant for our setting) dark soliton~\cite{Zakharov}.
For the coefficients corresponding to the Taylor expansion of the Hertz contact force, see Eq.~\eqref{eq:Taylor},
the NLS equation features a self-defocusing
nonlinearity. As such, the dark soliton is an exact solution of Eq.~\eqref{NLS}. Using the dark soliton
as the envelope function, we have the following
approximation,
\begin{equation} \label{ansatz2}
 y_n(t) = 2 \eps (-1)^{n+1}   \sqrt{\frac{\kappa}{\nu_3}} \tanh \left( \sqrt{ \frac{-\kappa}{ 2 \nu_2 }}   \epsilon (n - x_0)   \right) \cos( \omega_b t ) \qquad k_0=\pi, \, \omega_0 = 2 \sqrt{K_2/M}, \, c = 0
 \end{equation}
where $\omega_b = \omega_0 + \kappa \epsilon^2$ is the frequency of oscillation, $\kappa <0$ is 
a fixed but arbitrary parameter and $X_0 = \epsilon x_0$ is an arbitrary spatial translation; see~\cite{dark} for more details.

We seek to maximize how much the bead with the PZT element is squeezed: 
$$s_m(t):=u_{m-1} - u_{m+1} = y_m + y_{m+1}$$
where $m$ is the location of the PZT element. 
This quantity in the NLS approximation becomes:
 \begin{align}
 s_n(t) &= 2 \epsilon (-1)^{n} \sqrt{\frac{\kappa}{\nu_3}} \left( \tanh\left( \sqrt{\frac{-\kappa}{2\nu_2}} \epsilon (n+1-x_0) \right) - \tanh\left( \sqrt{\frac{-\kappa}{2\nu_2}} \epsilon (n - x_0) \right)    \right) \cos(\omega_b t)  \\
&\approx 2 \epsilon^2 (-1)^{n} \sqrt{\frac{\kappa}{\nu_3}} \left( 1 - \tanh^2\left( \sqrt{\frac{-\kappa}{2\nu_2}} \epsilon (n - x_0) \right)   \right) \cos(\omega_b t).
\end{align}
Thus, the profile of the squeeze is localized at the center of the chain, see Fig.~\ref{fig:NLS}(b). If we take $m$ to be at the center of the chain, and choose the actuation amplitude to correspond
to the saddle-node bifurcation (see e.g. Fig.~\ref{fig:NLS}(a)) we will have an analytical approximation, increasingly
valid in the limit of 
$ \tau \rightarrow 0$ and $a \rightarrow 0$ 
of how much the PZT bead is squeezed. To
compute the corresponding voltage, we can solve the following equation,
\begin{subequations}\label{eq:EMnls}
\begin{eqnarray} 
\frac{M}{2} \ddot{u}_m &= &\gamma [ \delta_{0} +   y_m ]_+^{3/2}  - K_a(\delta_1 +  u_m - u_{m+1} )  - d_{33} K_a V -   \frac{M}{2 \tau} \dot{u}_m ,    \hspace{.2cm}\\  
\frac{M}{2} \ddot{u}_{m+1} &=& K_a( \delta_1 + u_m - u_{m+1} ) -  \gamma [ \delta_{0} +   y_{m+1}]_+^{3/2}   + d_{33} K_a V -   \frac{M}{2 \tau} \dot{u}_{m+1} ,    \hspace{.2cm}\\  
R C(1 - k^2) \, \dot{V}  &=& - d_{33} K_a  R( \dot{u}_m - \dot{u}_{m+1}  ) -  V 
\end{eqnarray}
\end{subequations}
where $y_m ,y_{m+1}$ is given by Eq.~\eqref{ansatz2}. For out-of-phase actuation in a chain of odd length, we expect the solution
to be symmetric about the center bead, and hence we pick $x_0 = 1/2$ in Eq.~\eqref{ansatz2} and $m=0$ (for an infinite chain we assume $m=0$ is the center), which leads to,
$$ y_m(t) = y_{m+1}(t) =  \alpha \cos( \omega_b t ):= y(t), \qquad \alpha := 2 \epsilon   \sqrt{\frac{\kappa}{\nu_3}} \tanh \left( \sqrt{ \frac{-\kappa}{ 2 \nu_2 }}   \epsilon / 2   \right) $$
 Since $y = \OO(\epsilon) \ll \delta_0$, we expand the forcing terms in a Taylor series (see Eq.~\eqref{eq:Taylor}),
 %
and then we define $z = u_{m} - u_{m+1}$, which leads to
\begin{subequations}\label{eq:EMnls1}
\begin{eqnarray} 
\frac{M}{2} \ddot{z}  +  2 K_a   z   +   \frac{M}{2 \tau} \dot{z} + 2 d_{33} K_a V &= &   2( K_2 \, y(t) + K_3 \,y(t)^2 +  K_4 \, y(t)^3 ),    \hspace{.2cm}\\  
R C(1 - k^2) \, \dot{V}  &=& -d_{33} K_a  R \dot{z}  -  V 
\end{eqnarray}
\end{subequations}
which is a damped-driven linear ODE which can be solved exactly. While the full solution is cumbersome to express, the leading
order term to the steady-state amplitude of the voltage is easily found. Thus, we finally arrive at the following approximation
of the amplitude of the voltage,
\begin{subequations}\label{eq:Vmax}
\begin{eqnarray} 
V_{amp} &=& \frac{\alpha 4 K_2  \omega_b}{M \sqrt{(4K_a/M - \omega_b^2)^2 + (\omega_b/\tau)^2   } }\, \frac{d_{33} K_a R}{\sqrt{1 + ( CR\omega_b(1-k^2))^2}} \\
f_b &=&  \frac{ \omega_0 + \kappa \epsilon^2}{ 2 \pi }  \label{eq:Vmaxb} \\
a &=& \eps \sqrt{\frac{\kappa}{\nu_3}}.  \label{eq:Vmaxc}
\end{eqnarray}
\end{subequations}
%
To compute the driving amplitude $a$ we made use of Eq.~\eqref{ansatz2} and the approximation $|y_n| =  2|u_n|$ for $|n| \rightarrow \infty$.
Although the  approximation~\eqref{eq:Vmax} is only valid in the limit of zero damping, it performs fairly well using the experimentally measured dissipation constant of $\tau = 5\,\textrm{ms}$
for small values of $\omega_b - \omega_0=\epsilon^2 \kappa$ (i.e. for those close to the band edge), see Fig.~\ref{fig:NLS}(c). We note that this approximation can only be employed for the special combinations
of actuation frequency and drive given by Eqs.~\eqref{eq:Vmaxb}-\eqref{eq:Vmaxc} respectively.
Hence, it is evident that in the limit where the NLS approximation becomes
relevant, the voltage between the electrodes of the PZT can not only
be numerically computed but also
analytically approximated.

\section{Conclusions and Future Directions} \label{sec:theend}

We considered a harmonically driven 1D granular chain with an embedded piezoelectric sensor (a PZT) at the center of the chain.  
Vibrational states of the nonlinear system,
which were identified by measuring the voltage output of the PZT sensor, were found to exist for much broader range of input frequencies
when compared to the linearized counterpart of this system. The electromechanical model we derived provided good agreement
to the experimentally observed values under both static and dynamic driving frequency, although it may be less accurate for large amplitude (more highly
nonlinear) states. 
The electromechanical model provides an insightful description
of how the electrical system can affect the dynamics of the mechanical system, and could be useful for future studies in granular chains that
utilize PZT sensors to deduce force, without relying on calibration factors. The NLS equation also provided an analytical
prediction of the voltage production for certain combinations of driving frequency and amplitude that compared favorably to
numerical simulations.
While this study focused on out-of-phase boundary excitations, 
we suspect that other harmonic excitation conditions (where e.g. the 
phase difference
between the two driving frequencies is arbitrary) will yield qualitatively 
similar results, although the optimal location for the PZT may no
longer be the center. Thus, an interesting future direction would be to study a chain in which different particles contain a PZT and the chain is subject to
various driving boundary excitations (not necessarily out-of-phase). 
Optimizing the relevant mechanism and identifying the relation between
optimal voltage output at different sites as a function of phase would
certainly be of interest to future work.

We believe this study lays a foundation for future investigations of energy harvesting in spatially extended nonlinear dynamical systems, identifying
their capability for more broadband harvesting, but also raising some
of their limitations including the scale of the phenomenon and the partial
lack of (numerical) control of the output of the mechanism for larger
amplitudes/high nonlinearities.
While the granular chain itself may not be ideal for energy harvesting purposes, it demonstrates that the combination
of multiple modes and nonlinearity can lead to the possibility of 
different vibrational  states (including high energy ones)
for a broad range of input frequencies.


\begin{acknowledgments}
C.C. was partially supported by the ETH Zurich Foundation through the Seed Project ESC-A 06-14. 
E.G.C, C.D. and P.G.K acknowledge support from the US-AFOSR under grant FA9550-12-10332.
P.G.K. also acknowledges support from the NSF under grant DMS-1312856, from ERC and
FP7-People under grant 605096, and from the Binational (US-Israel) Science Foundation
through grant 2010239. P.G.K.'s work at Los Alamos is supported in part by the U.S. 
Department of Energy. J.Y. thanks the support of the NSF (CMMI-1414748) and the US-ONR (N000141410388).
\end{acknowledgments}

\end{document}